\documentclass[prl,showpacs,twocolumn,floatfix]{revtex4}
\usepackage{graphicx}

\begin{document}
\title{Nonlinear cascades in two-dimensional turbulent magnetoconvection}
\author{Dan \v{S}kandera}
\author{Wolf-Christian M\"{u}ller}
\email{Wolf.Mueller@ipp.mpg.de}
\affiliation{Max-Planck-Institut f\"{u}r Plasmaphysik, 85748 Garching, Germany}
\date{\today}

\begin{abstract}
The dynamics of spectral transport in two-dimensional turbulent convection of 
electrically conducting fluids is studied by means of direct numerical simulations (DNS) 
in the frame of the magnetohydrodynamic (MHD) Boussinesq approximation. 
The system performs 
quasi-oscillations between two different regimes of small-scale turbulence: one 
dominated by nonlinear MHD interactions,
the other governed by buoyancy forces. The self-excited change of turbulent 
states is reported here for the first time.
The process is controlled by the ideal invariant cross-helicity, $H^\mathrm{C}=\int_S \mathrm{d}S \mathbf{v}\cdot\mathbf{b}$.
The observations are explained by the interplay of convective driving with
the nonlinear spectral transfer of total MHD energy and cross-helicity.
\end{abstract}

\maketitle

Turbulent convection of an electrically conducting fluid is of major 
importance for the dynamics of
stellar convection zones and the evolution of magnetic fields in these regions.
It is thus necessary to better understand its nonlinear dynamics which can be strongly influenced
by advected temperature fluctuations and self-organization processes associated with the 
magnetic field. 
To this end the two-dimensional magnetohydrodynamic (MHD) Boussinesq approximation is applied
as a simplified model 
that asymptotically describes the behavior of plasmas under the influence 
of a strong mean magnetic field directed perpendicular to the direction of
gravity --- a configuration for example realized in the filamentary sunspot penumbra 
\cite{thomas_weiss:sunspots} where the magnetic field is nearly horizontal and in 
laboratory plasma experiments. Additionally,
convection represents a natural and more realistic way of 
sustaining turbulence when studying its inherent properties by numerical simulations in contrast to 
the somewhat artificial forcing mechanisms that are often applied in such studies. 
Buoyancy can turn the temperature field into an
active scalar and, as a consequence, substantially modify the nonlinear 
spectral transport of energy and other ideal invariants by the
respective turbulent cascade.  Some progress has been made with regard
to the understanding of \textit{homogeneous} hydrodynamic turbulent
convection,
e.g. \cite{lvov:scalings,grossmann_lvov:scalings,procaccia_zeitak:convscal,brandenburg:convecshell,suzuki_toh:convecsim2D,suzuki_toh:convecshell,biskamp_schwartz:convecsim2Da,celani_etal:convecsim2D,calzavarini_etal:elevator}.
However, studies of convection in magnetofluids (see, e.g.,
\cite{proctor:mconvecreview} for a recent review) which focus on the
small-scale properties of homogeneously turbulent states remain scarce.

This Letter deals with an investigation of the inertial-range dynamics of 
convectively driven two-dimensional
MHD turbulence by direct numerical simulation (DNS).  
The system displays quasi-oscillatory changes between a state where
turbulence is dominated by buoyant forces and a state
governed by nonlinear MHD interactions.  The constantly changing
cross-helicity, $H^\mathrm{C}=\int_S\mathrm{d}S\mathbf{v}\cdot\mathbf{b}$,  is the cause for this
behavior since it regulates the importance of MHD nonlinearities
compared to buoyancy effects. The nonlinear interaction of the turbulent cascades of energy 
and cross-helicity which leads to the self-excited alternation of turbulent regimes is 
reported here for the first time and yields further insight into the yet not fully 
understood dynamics of turbulent flows.
    
The system is described by the Boussinesq MHD equations, see for
example \cite{proctor:mconvecreview}. In two dimensions with $x$
denoting the horizontal and $z$ the vertical direction they read
\begin{eqnarray} \frac{\partial\omega}{\partial
t}+\mathbf{v}\cdot\nabla\omega-\mathbf{b}\cdot\nabla j&=&
-\partial_x\theta+\nu\Delta\omega,\label{eqn:Bouss1}\\
\frac{\partial\psi}{\partial
t}+(\mathbf{v}\cdot\nabla)\psi&=&\eta\Delta\psi,\label{eqn:Bouss2}\\
\frac{\partial\theta}{\partial
t}+(\mathbf{v}\cdot\nabla)\theta&=&v_x+\kappa\Delta\theta,\label{eqn:Bouss3}\\
\nabla\cdot\mathbf{v}=\nabla\cdot\mathbf{b}&=&0\label{eqn:Bouss4}
\end{eqnarray} where $\theta$ denotes temperature fluctuations about a
mean gradient, $\mathbf{v}$ is the velocity of the flow,
$\mathbf{b}$ is the magnetic field, $\omega=\mathbf{e}_y\cdot(\nabla\times\mathbf{v})$ stands
for the vorticity and $\psi$ for the scalar
magnetic potential, $\mathbf{b}=\nabla\psi\times\mathbf{e}_y$. The symbol $\mathbf{e}_y$
denotes the unit normal vector of the two-dimensional plane.  The
current density is given by $j=-\Delta\psi$. The
dissipation coefficients $\nu$, $\eta$, $\kappa$ are dimensionless
kinematic viscosity, resistivity and thermal diffusivity,
respectively. The equations are given in non-dimensional form using
a normalization to the time characteristic of large-scale
buoyant motions $t_\mathrm{b}=(\alpha g |\nabla T_0|)^{-1/2}$ and the
temperature gradient scale $L_0=T_*/|\nabla T_0|$. Here, $\alpha$ is
the coefficient of thermal expansion, $g$ is the gravitational
acceleration acting in the negative $z$-direction, and $T_0(x)$ is the
mean linear temperature profile. The relevant definition for 
$T_*$ is the root-mean-square (RMS) value of temperature fluctuations, see e.g. \cite{gibert_etal:T0}.
In contrast to the usual setup, $\nabla T_0$ 
is in the horizontal direction in order to eliminate
the elevator instability \cite{calzavarini_etal:elevator}.  It otherwise appears due
to periodic boundary conditions in the $z$-direction and leads to
the formation of coherent vertical jets that significantly degrade the
quality of turbulence statistics. 
A horizontal $\nabla T_0$ leads to  a large-scale $x$-dependent buoyancy force that
can not be balanced by the pressure gradient.  Since this work
concentrates on the small-scale dynamics of turbulence the generation
of large-scale vorticity by this effect is neglected in
Eq. (\ref{eqn:Bouss1}).  The mean temperature gradient drives the turbulent flow 
(right-hand-side (RHS) of Eq. (\ref{eqn:Bouss3}), first term) by temperature fluctuations
which couple with 
velocity fluctuations through the buoyancy force
(RHS of Eq. (\ref{eqn:Bouss1}), first term).  The absence
of magnetic dynamo action in 2D MHD \cite{zeldovich:antidynamo2D,Cow55} would
necessitate explicit driving of magnetic field fluctuations. However as the flow 
is homogeneously turbulent and thus does not concentrate the magnetic field 
on the boundaries of convection cells,
the decay of magnetic energy is a slow process on the resistive time
scale $t_\mathrm{D}=L_0^2/\eta\gg t_\mathrm{b}$. Thus, this
quantity is quasi-stationary during the simulation 
which extends over $15$ large-scale buoyancy times
$t_\mathrm{b}$.

The set of equations (\ref{eqn:Bouss1})--(\ref{eqn:Bouss4}) is solved
on a $2\pi$-biperiodic square using a standard pseudospectral method
with dealiasing according to the 2/3 rule \cite{canuto:book}.  A
simulation of the above-described setup is conducted with
a resolution of $2048^2$ collocation points. The initial
state consists of  randomly generated fields.
The parameters of the run are set to
$\nu=\eta=7\times 10^{-4}$ and $\kappa=1.3\times 10^{-4}$ which
corresponds to a Prandtl number $\mathsf{Pr}=\nu/\kappa\approx 5.4$
and a magnetic Prandtl number $\mathsf{Pr_m}=\nu/\eta=1$. The nominal
Rayleigh number is $\mathsf{Ra}\approx 2\times 10^6$.
We note, however, that the value of $\mathsf{Ra}$  
characteristic of Rayleigh-B\'enard systems with impermeable vertical boundaries  
is only of limited significance for the present periodic setup.
All parameters are chosen such as to obtain comparable wave-number
intervals for the inertial ranges of the turbulent fields. The
inherent anisotropy of the flow mainly affects the
largest-scale fluctuations with wavenumbers $k\leq 4$, while at smaller scales 
anisotropy becomes negligible.  The total energy is given by
$E=E^\mathrm{mhd}+E^\theta$ with 
$E^\mathrm{mhd}=E^\mathrm{k}+E^\mathrm{m}=1/2\int_S\mathrm{d}S(v^2+b^2)$
and the temperature ``energy''
$E^\theta=1/2\int_S\mathrm{d}S\theta^2$. Analogously, the symbol
$\varepsilon=\varepsilon_\mathrm{mhd}+\varepsilon_\theta=\int_S\mathrm{d}S
\left[(\mu\omega^2+\eta j^2)+\kappa(\nabla\theta)^2\right]$ denotes
the sum of dissipation rates of total magnetohydrodynamic energy and
temperature energy. The turbulent state which develops quickly after the onset of convective instability
is initially characterized by $E^\mathrm{k}\approx 5.0$, $E^\mathrm{m}\approx 6.0$, and $E^\theta\approx 1.9$.
The values increase slowly by about $40\%$ over the simulation period while staying in roughly constant ratios to each other.

\begin{figure}
\includegraphics[width=7.6cm]{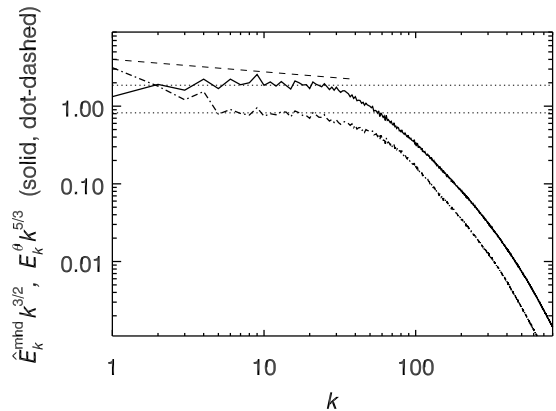} \caption{\label{fig1}
Angle-integrated spectra of normalized MHD energy,
$\hat{E}_k^\mathrm{mhd}=E^\mathrm{mhd}_k/(b_0\varepsilon_\mathrm{mhd})^{1/2}$, compensated by
$k^{3/2}$ (solid line), and temperature ``energy'', $E^\theta_k$,
compensated by $k^{5/3}$ (dash-dotted line), time-averaged over the IK-intervals. 
The dashed line indicates $k^{-5/3}$-scaling
with regard to $\hat{E}_k^\mathrm{mhd}$, the dotted horizontals mark scaling law prefactors.}
\end{figure}

In the observed energy spectra the signatures of two distinct turbulent regimes appear which
quasi-periodically alternate. This happens in a continuous transition where either one
or the other signature becomes more dominant. The time characteristic of the alternations and the duration 
of the clearest manifestation of the different states is of the order of $t_\mathrm{b}$.
The simulation period consists of two
groups of time-intervals labeled ``BO'' and ``IK''.  Angle-integrated
energy spectra, $E_k=\int \mathrm{d}^3k'\delta(|\mathbf{k'}|-k)E(\mathbf{k'})$, obtained by time-averaging over the group of IK
intervals are shown in Fig.~\ref{fig1}.  
The spectrum of total MHD energy denoted by the solid line exhibits
Iroshnikov-Kraichnan (IK) behavior,
$E^\mathrm{mhd}_k\sim k^{-3/2}$ \cite{iroshnikov:ikmodel,kraichnan:ikmodel}, in the inertial
range of scales, $5\alt k\alt 20$.  The spectrum is normalized by 
$(\varepsilon_\mathrm{mhd}b_0)^{1/2}$ with $b_0$ being the RMS
magnetic field and it displays a constant prefactor $C_\mathrm{IK}\approx 1.8$ that
agrees well with previous work on 2D MHD turbulence, see, e.g.,
\cite{biskamp_schwarz:2Dmhd}. The spectra of $E^k$ and $E^m$ (not shown) exhibit 
comparable amplitudes and scaling consistent with the IK picture. 
Although the validity of this 2D MHD turbulence phenomenology 
is still a matter of controversy due to its neglect of anisotropy 
caused by the magnetic field, DNS of two-dimensional 
MHD turbulence
\cite{biskamp_welter:2dmhddecay,biskamp_schwarz:2Dmhd} appear to agree well with it.

The temperature energy spectrum $E^\theta_k$ displays 
a somewhat shorter scaling range, $5\alt
k\alt 12$, with an exponent of $-5/3$. The Obukhov-Corrsin 
scaling suggests that temperature fluctuations are 
passively advected by the velocity field.  
Note that this does not necessarily imply the same scaling for kinetic energy which 
in fact scales closer to $k^{-3/2}$.
The spectrum of the
mean square magnetic potential (not shown)  scales self-similarly
over the same spectral interval as $E_k^\mathrm{mhd}$ with $
|\psi_k|^2\sim k^{-7/2}$, consistent
with magnetic field fluctuations under the influence of large-scale
driving, $|\psi_k|^2\sim E^\mathrm{mhd}_kk^{-2}$ \cite{pouquet:invcasc2D}. 
Thus, during the IK phases the effects of buoyancy on the
flow are negligible compared to the MHD nonlinearities which clearly
dominate nonlinear dynamics.

\begin{figure}
\includegraphics[width=7.6cm]{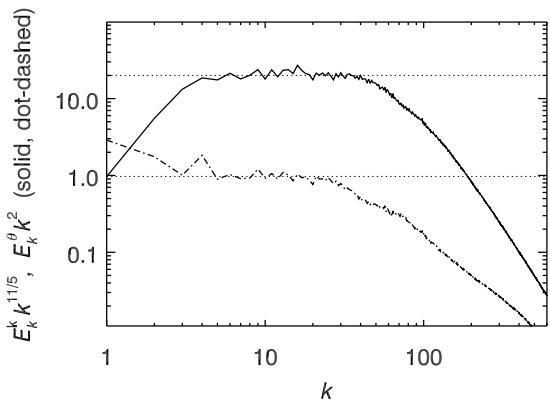}
\caption{\label{fig2}Angle-integrated spectra of kinetic energy,
$E^\mathrm{k}_k$, compensated by $k^{11/5}$ (solid line), and temperature
energy, $E^\theta_k$, compensated by $k^{2}$ (dashed-dotted
line). Both spectra are time-averaged over the BO-intervals of the
simulation. Same meaning of dotted lines as in Fig. \ref{fig1}.}  \end{figure}

Angle-integrated energy spectra obtained by time-averaging over the
``BO'' intervals are depicted in Fig.~\ref{fig2}.   
Although
$E_k^\mathrm{k}$ (solid line)  is not an ideal invariant of MHD 
the spectrum scales with the 
exponent $-11/5$ for $4\alt k\alt 40$ 
while for the temperature energy (dash-dotted line) approximately
$E^\theta_k\sim k^{-2}$ is obtained.  
No clear scaling behavior is exhibited by $E^\mathrm{mhd}_k$ and $E_k^\mathrm{m}$. 
In the Bolgiano-Obukhov (BO) picture
of convective turbulence \cite{bolgiano:convecpheno,obukhov:convecpheno} the temperature field can drive velocity
fluctuations. 
As opposed to hydrodynamic turbulence, spectral transfer is not characterized by the turnover time at
scale $\ell$, $t_\mathrm{NL}\sim \ell/v_\ell$, but by the
scale-dependent buoyancy time, $t_b^*\sim
(\ell/\theta_\ell)^{1/2}$. Consequently, $E^\theta_k\sim k^{-7/5}$ and
$E^\mathrm{k}_k\sim k^{-11/5}$. The observed scaling of $E^k_k$ is in good agreement with the respective BO-law, in contrast to the behaviour of
$E^\theta_{k}$. Hydrodynamic test runs with similar parameters exhibit BO-behaviour for
both fields.  The
spectra shown in Fig.~\ref{fig2} suggest that the investigated system
operates in a modified buoyancy-dominated Bolgiano-Obukhov-like 
regime. The buoyancy force
dominates nonlinear MHD interactions in the inertial range, i.e. the
temperature is an active scalar significantly influencing the
energy transfer.  

\begin{figure}
\includegraphics[width=7.6cm]{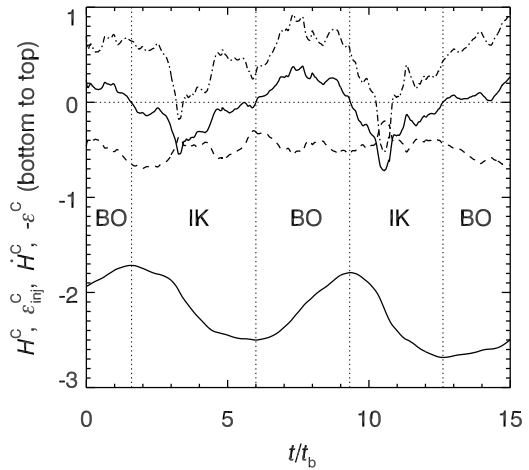}
\caption{\label{fig4}Time-evolution of $H^\mathrm{C}$ (lower solid line), $\dot{H}^\mathrm{C}$ (upper solid line), 
the associated flux injected by convective driving $\varepsilon^\mathrm{C}_\mathrm{inj}$ (dash-dotted line)
and the dissipation rate $\varepsilon^\mathrm{C}$ (dashed line).
The intervals of IK- and BO-states are indicated.}
\end{figure}

The time intervals of both turbulent regimes are correlated with 
the evolution of total cross-helicity $H^\mathrm{C}$ 
as depicted in  Fig.~\ref{fig4}. 
It is important to note that other physical quantities, e.g. the energies or the turbulent 
heat flux, fluctuate on different time scales and are not correlated with the quasi-oscillations. 
A simulation run with lower resolution $1024^2$ confirms the stability of the
phenomenon for at least $100 t_b$.  The dissipation coefficients of this test run
are set to $\nu=1.5\times 10^{-3}$, $\eta=7.5\times 10^{-4}$,
$\kappa=4\times 10^{-4}$. The impact of cross-helicity in a strongly aligned turbulent system 
like in this simulation with  
$\rho=\langle|\mathbf{v}\cdot\mathbf{b}|\rangle/(4E^\mathrm{k}E^\mathrm{m})^{1/2}\in [0.8,0.95]$ 
on the nonlinear dynamics of MHD turbulence is known to lead to significant modifications of 
spectral energy transfer \cite{grappin_etal:zpzmspectra,grappin_pouquet_leorat:correl}.
The limiting case $\rho=1$ 
corresponds to an Alfv\'enic state where the nonlinear dynamics is completely switched off. 
The effect of weakening of nonlinear MHD interactions in high cross-helicity states of 
turbulence is the key ingredient in following considerations.

In contrast to plain MHD turbulence, cross-helicity in magnetoconvective turbulence is not 
an ideal invariant as can be seen from $\dot{H}^\mathrm{C}=\int_S\theta b_z dS-(\nu+\eta)\int_S j\omega dS=\varepsilon^\mathrm{C}_\mathrm{inj}-\varepsilon^\mathrm{C}$.
The dissipative term $\varepsilon^\mathrm{C}$ has 
a larger effect at small scales due to the appearance of spatial derivatives whereas the term 
$\varepsilon^\mathrm{C}_\mathrm{inj}$ injects cross-helicity 
predominantly at large scales.
The interplay between all three terms of the cross-helicity balance in the performed simulation is 
shown in Fig.~\ref{fig4}. 
The time intervals of IK turbulence can be identified with negative
$\dot{H}^\mathrm{C}$ (upper solid line) whereas the time
intervals of BO-like turbulence are correlated with positive
$\dot{H}^\mathrm{C}$. The term $\varepsilon_\mathrm{inj}^\mathrm{C}$ (dashed line) is always
negative permanently injecting negative cross-helicity into the
system. Fig.~\ref{fig4} shows that if the small-scale
cross-helicity dissipation $\varepsilon^\mathrm{C}$ (dot-dashed line) 
becomes larger than the absolute amount of the
injected cross-helicity, $|\varepsilon^\mathrm{C}_\mathrm{inj}|$, the system
switches from the IK state to the BO state and vice versa.

\begin{figure}
\includegraphics[width=7.6cm,height=6cm]{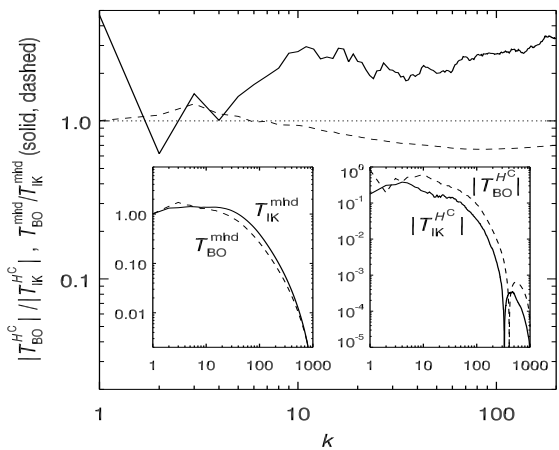}
\caption{\label{fig3}Ratio of nonlinear spectral energy fluxes of MHD energy $T_k^\mathrm{mhd}=\int_k^{k_\mathrm{max}}\mathrm{dk'}\{\mathbf{v}\cdot[-\mathbf{v}\cdot\nabla\mathbf{v}+\mathbf{b}\cdot\nabla\mathbf{b}-\nabla P]+\mathbf{b}\cdot[\nabla\times(\mathbf{v}\times\mathbf{b})]\}_{k'}$ (dashed line) 
and cross-helicity $T_k^{H^\mathrm{C}}=T_k^+-T_k^-=1/2\int_k^{k_\mathrm{max}}\mathrm{d}k'\{-\mathbf{z}^+\cdot[\mathbf{z^-\cdot\nabla\mathbf{z^+}}+\nabla P]+\mathbf{z}^-\cdot[\mathbf{z}^+\cdot\nabla\mathbf{z}^-+\nabla P]\}_{k'}$ (solid line) time-averaged over 
IK- and BO-intervals with the total pressure $P=p+b^2/2$ and  $\{\bullet\}_{k}$ denoting Fourier transformation. Insets display the spectral fluxes with  $T_k^{H^C}>0$ for $k\lesssim 300$ and negative beyond.}
\end{figure}

In the following, an explanation for the quasi-oscillations is
proposed.  It is assumed that at the beginning the system operates in
the IK regime. This choice is not essential, but provides a
convenient starting point for the ensuing considerations. The IK
regime implies that inertial-range energies and energy fluxes associated with Els\"asser variables
$\mathbf{z}^\pm=\mathbf{v}\pm\mathbf{b}$ are approximately equal
\cite{dobrowolny:ik}, i.e.  $E^+\approx E^-$ with
$E^\pm=1/4\int_S\mathrm{d}S(z^\pm)^2$ and for the corresponding
direct nonlinear spectral fluxes $T_k^+\approx T_k^-$.
Consequently, total cross-helicity, $H^\mathrm{C}=E^+-E^-$, and
the associated nonlinear flux in the inertial range,
$T_k^{H^\mathrm{C}}=T^+_k-T^-_k\approx\varepsilon^+-\varepsilon^-$, are minimal.
Nonlinear MHD interactions are
dominant in the inertial-range (see. Fig.~\ref{fig3}). However, the convective
driving continuously generates negative cross-helicity at the rate
$\varepsilon^\mathrm{C}_\mathrm{inj}$, predominantly at
largest-scales.
The apparently delayed adaption of the initially depleted direct nonlinear transfer of cross-helicity  leads to an
accumulation of $H^\mathrm{C}$ at large scales.
The resulting growth of the $|H^\mathrm{C}|$ breaks the balance of $E^+$ and $E^-$  
and, consequently, weakens nonlinear MHD interactions together with the resulting 
spectral flux of $E^\mathrm{mhd}$ as shown in Fig.~\ref{fig3}.
Thus, the process gives rise to inertial-range dynamics dominated by
buoyancy forces. Therefore, simultaneously with the growth of
 $H^\mathrm{C}$ the dynamics of the system changes toward the
buoyancy dominated BO-like regime of turbulence.

In the BO-like regime, due to $E^+\neq E^-$, the corresponding spectral
fluxes $T_k^+$ and $T_k^-$ are different. Hence, 
the spectral flux $T_k^{H^\mathrm{C}}$ and cross-helicity dissipation $\varepsilon^\mathrm{C}$ 
are larger. 
Thus,
the BO-like regime leads to efficient annihilation of the accumulated
cross-helicity. As $H^\mathrm{C}$ decreases, the system 
approaches Els\"asser energy equipartition, $E^+\approx E^-$, and it
returns to the IK regime of 2D magnetoconvective turbulence.
Due to the continuous injection of cross-helicity 
by large-scale convection, magnetic and velocity field of the flow are
highly aligned.  The same phenomenon could not be observed in topologically less constrained
three-dimensional simulations. 

In summary, self-excited quasi-oscillations between an
Iroshnikov-Kraichnan regime and a Bolgiano-Obukhov-like regime of turbulence are
observed in two-dimensional DNS of
homogeneous MHD Boussinesq turbulence.  The highly aligned turbulent
fields, a consequence of the convective large-scale driving, give
rise to a quasi-periodic weakening of nonlinear MHD interactions in
favor of buoyancy effects. The resulting Bolgiano-Obukhov-like regime allows the removal of
cross-helicity by nonlinear spectral transfer strengthening the
MHD nonlinearities again.  The two distinct turbulent states emerge due to the nonlinear 
interplay of the cascades of energy and cross-helicity, a situation of general importance to
MHD systems constrained by the presence of a strong mean magnetic field as found, e.g.,
on the Sun.  

\begin{acknowledgments}
The authors thank M. Proctor for an important remark regarding the simulation setup.     
\end{acknowledgments}
%

\begin{thebibliography}{25}
\expandafter\ifx\csname natexlab\endcsname\relax\def\natexlab#1{#1}\fi
\expandafter\ifx\csname bibnamefont\endcsname\relax
  \def\bibnamefont#1{#1}\fi
\expandafter\ifx\csname bibfnamefont\endcsname\relax
  \def\bibfnamefont#1{#1}\fi
\expandafter\ifx\csname citenamefont\endcsname\relax
  \def\citenamefont#1{#1}\fi
\expandafter\ifx\csname url\endcsname\relax
  \def\url#1{\texttt{#1}}\fi
\expandafter\ifx\csname urlprefix\endcsname\relax\def\urlprefix{URL }\fi
\providecommand{\bibinfo}[2]{#2}
\providecommand{\eprint}[2][]{\url{#2}}

\bibitem[{\citenamefont{Thomas and Weiss}(2004)}]{thomas_weiss:sunspots}
\bibinfo{author}{\bibfnamefont{J.~H.} \bibnamefont{Thomas}} \bibnamefont{and}
  \bibinfo{author}{\bibfnamefont{N.~O.} \bibnamefont{Weiss}},
  \bibinfo{journal}{Annual Review of Astronomy and Astrophysics}
  \textbf{\bibinfo{volume}{42}}, \bibinfo{pages}{517} (\bibinfo{year}{2004}).

\bibitem[{\citenamefont{Procaccia and
  Zeitak}(1989)}]{procaccia_zeitak:convscal}
\bibinfo{author}{\bibfnamefont{I.}~\bibnamefont{Procaccia}} \bibnamefont{and}
  \bibinfo{author}{\bibfnamefont{R.}~\bibnamefont{Zeitak}},
  \bibinfo{journal}{Physical Review Letters} \textbf{\bibinfo{volume}{62}},
  \bibinfo{pages}{2128} (\bibinfo{year}{1989}).

\bibitem[{\citenamefont{Brandenburg}(1992)}]{brandenburg:convecshell}
\bibinfo{author}{\bibfnamefont{A.}~\bibnamefont{Brandenburg}},
  \bibinfo{journal}{Physical Review Letters} \textbf{\bibinfo{volume}{69}},
  \bibinfo{pages}{605} (\bibinfo{year}{1992}).

\bibitem[{\citenamefont{Toh and Suzuki}(1994)}]{suzuki_toh:convecsim2D}
\bibinfo{author}{\bibfnamefont{S.}~\bibnamefont{Toh}} \bibnamefont{and}
  \bibinfo{author}{\bibfnamefont{E.}~\bibnamefont{Suzuki}},
  \bibinfo{journal}{Physical Review Letters} \textbf{\bibinfo{volume}{73}},
  \bibinfo{pages}{1501} (\bibinfo{year}{1994}).


\bibitem[{\citenamefont{Suzuki and Toh}(1995)}]{suzuki_toh:convecshell}
\bibinfo{author}{\bibfnamefont{E.}~\bibnamefont{Suzuki}} \bibnamefont{and}
  \bibinfo{author}{\bibfnamefont{S.}~\bibnamefont{Toh}},
  \bibinfo{journal}{Physical Review E} \textbf{\bibinfo{volume}{51}},
  \bibinfo{pages}{5628} (\bibinfo{year}{1995}).

\bibitem[{\citenamefont{Biskamp and
  Schwarz}(1997)}]{biskamp_schwartz:convecsim2Da}
\bibinfo{author}{\bibfnamefont{D.}~\bibnamefont{Biskamp}} \bibnamefont{and}
  \bibinfo{author}{\bibfnamefont{E.}~\bibnamefont{Schwarz}},
  \bibinfo{journal}{Europhysics Letters} \textbf{\bibinfo{volume}{40}},
  \bibinfo{pages}{637} (\bibinfo{year}{1997}).

\bibitem[{\citenamefont{L'vov}(1991)}]{lvov:scalings}
\bibinfo{author}{\bibfnamefont{V.~S.} \bibnamefont{L'vov}},
  \bibinfo{journal}{Physical Review Letters} \textbf{\bibinfo{volume}{67}},
  \bibinfo{pages}{687} (\bibinfo{year}{1991}).

\bibitem[{\citenamefont{Grossmann and L'vov}(1993)}]{grossmann_lvov:scalings}
\bibinfo{author}{\bibfnamefont{S.}~\bibnamefont{Grossmann}} \bibnamefont{and}
  \bibinfo{author}{\bibfnamefont{V.~S.} \bibnamefont{L'vov}},
  \bibinfo{journal}{Physical Review E} \textbf{\bibinfo{volume}{47}},
  \bibinfo{pages}{4161} (\bibinfo{year}{1993}).

\bibitem[{\citenamefont{Celani et~al.}(2002)\citenamefont{Celani, Matsumoto,
  Mazzino, and Vergassola}}]{celani_etal:convecsim2D}
\bibinfo{author}{\bibfnamefont{A.}~\bibnamefont{Celani}},
  \bibinfo{author}{\bibfnamefont{T.}~\bibnamefont{Matsumoto}},
  \bibinfo{author}{\bibfnamefont{A.}~\bibnamefont{Mazzino}}, \bibnamefont{and}
  \bibinfo{author}{\bibfnamefont{M.}~\bibnamefont{Vergassola}},
  \bibinfo{journal}{Physical Review Letters} \textbf{\bibinfo{volume}{88}},
  \bibinfo{pages}{054503} (\bibinfo{year}{2002}).

\bibitem[{\citenamefont{Calzavarini et~al.}(2005)\citenamefont{Calzavarini,
  Lohse, Toschi, and Tripiccione}}]{calzavarini_etal:elevator}
\bibinfo{author}{\bibfnamefont{E.}~\bibnamefont{Calzavarini}},
  \bibinfo{author}{\bibfnamefont{D.}~\bibnamefont{Lohse}},
  \bibinfo{author}{\bibfnamefont{F.}~\bibnamefont{Toschi}}, \bibnamefont{and}
  \bibinfo{author}{\bibfnamefont{R.}~\bibnamefont{Tripiccione}},
  \bibinfo{journal}{Physics of Fluids} \textbf{\bibinfo{volume}{17}},
  \bibinfo{pages}{055107} (\bibinfo{year}{2005}).

\bibitem[{\citenamefont{Proctor}(2005)}]{proctor:mconvecreview}
\bibinfo{author}{\bibfnamefont{M.~R.~E.} \bibnamefont{Proctor}}, in
  \emph{\bibinfo{booktitle}{Fluid Dynamics and Dynamos in Astrophysics and
  Geophysics}}, edited by \bibinfo{editor}{\bibfnamefont{A.~M.}
  \bibnamefont{Soward}}, \bibinfo{editor}{\bibfnamefont{C.~A.}
  \bibnamefont{Jones}}, \bibinfo{editor}{\bibfnamefont{D.~W.}
  \bibnamefont{Hughes}}, \bibnamefont{and}
  \bibinfo{editor}{\bibfnamefont{N.~O.} \bibnamefont{Weiss}}
  (\bibinfo{publisher}{CRC Press}, \bibinfo{address}{Boca Raton, Florida},
  \bibinfo{year}{2005}), pp. \bibinfo{pages}{235--276}.

\bibitem[{\citenamefont{Gibert et~al.}(2006)\citenamefont{Gibert, Pabiou,
  Chill\`a, and Castaing}}]{gibert_etal:T0}
\bibinfo{author}{\bibfnamefont{M.}~\bibnamefont{Gibert}},
  \bibinfo{author}{\bibfnamefont{H.}~\bibnamefont{Pabiou}},
  \bibinfo{author}{\bibfnamefont{F.}~\bibnamefont{Chill\`a}}, \bibnamefont{and}
  \bibinfo{author}{\bibfnamefont{B.}~\bibnamefont{Castaing}},
  \bibinfo{journal}{Physical Review Letters} \textbf{\bibinfo{volume}{96}},
  \bibinfo{pages}{084501} (\bibinfo{year}{2006}).

\bibitem[{\citenamefont{Zeldovich}(1957)}]{zeldovich:antidynamo2D}
\bibinfo{author}{\bibfnamefont{Y.~B.} \bibnamefont{Zeldovich}},
  \bibinfo{journal}{Soviet Physics JETP} \textbf{\bibinfo{volume}{4}},
  \bibinfo{pages}{460} (\bibinfo{year}{1957}).

\bibitem[{\citenamefont{Cowling}(1955)}]{Cow55}
\bibinfo{author}{\bibfnamefont{T.~G.} \bibnamefont{Cowling}},
  \bibinfo{journal}{Vistas in Astronomy} \textbf{\bibinfo{volume}{1}},
  \bibinfo{pages}{313} (\bibinfo{year}{1955}).

\bibitem[{\citenamefont{Canuto et~al.}(1988)\citenamefont{Canuto, Hussaini,
  Quarteroni, and Zang}}]{canuto:book}
\bibinfo{author}{\bibfnamefont{C.}~\bibnamefont{Canuto}},
  \bibinfo{author}{\bibfnamefont{M.~Y.} \bibnamefont{Hussaini}},
  \bibinfo{author}{\bibfnamefont{A.}~\bibnamefont{Quarteroni}},
  \bibnamefont{and} \bibinfo{author}{\bibfnamefont{T.~A.} \bibnamefont{Zang}},
  \emph{\bibinfo{title}{Spectral Methods in Fluid Dynamics}}
  (\bibinfo{publisher}{Springer-Verlag}, \bibinfo{address}{New York},
  \bibinfo{year}{1988}).

\bibitem[{\citenamefont{Iroshnikov}(1964)}]{iroshnikov:ikmodel}
\bibinfo{author}{\bibfnamefont{P.~S.} \bibnamefont{Iroshnikov}},
  \bibinfo{journal}{Soviet Astronomy} \textbf{\bibinfo{volume}{7}},
  \bibinfo{pages}{566} (\bibinfo{year}{1964}), \bibinfo{note}{[Astron. Zh.,
  {40}:742, 1963]}.

\bibitem[{\citenamefont{Kraichnan}(1965)}]{kraichnan:ikmodel}
\bibinfo{author}{\bibfnamefont{R.~H.} \bibnamefont{Kraichnan}},
  \bibinfo{journal}{Physics of Fluids} \textbf{\bibinfo{volume}{8}},
  \bibinfo{pages}{1385} (\bibinfo{year}{1965}).

\bibitem[{\citenamefont{Biskamp and Schwarz}(2001)}]{biskamp_schwarz:2Dmhd}
\bibinfo{author}{\bibfnamefont{D.}~\bibnamefont{Biskamp}} \bibnamefont{and}
  \bibinfo{author}{\bibfnamefont{E.}~\bibnamefont{Schwarz}},
  \bibinfo{journal}{Physics of Plasmas} \textbf{\bibinfo{volume}{8}},
  \bibinfo{pages}{3282} (\bibinfo{year}{2001}).

\bibitem[{\citenamefont{Biskamp{ \nop{g}und H.
  Welter}}(1989)}]{biskamp_welter:2dmhddecay}
\bibinfo{author}{\bibfnamefont{D.}~\bibnamefont{Biskamp{ \nop{g}und H.
  Welter}}}, \bibinfo{journal}{Physics of Fluids B}
  \textbf{\bibinfo{volume}{1}}, \bibinfo{pages}{1964} (\bibinfo{year}{1989}).

\bibitem[{\citenamefont{Pouquet}(1978)}]{pouquet:invcasc2D}
\bibinfo{author}{\bibfnamefont{A.}~\bibnamefont{Pouquet}},
  \bibinfo{journal}{Journal of Fluid Mechanics} \textbf{\bibinfo{volume}{88}},
  \bibinfo{pages}{1} (\bibinfo{year}{1978}).

\bibitem[{\citenamefont{Bolgiano}(1959)}]{bolgiano:convecpheno}
\bibinfo{author}{\bibfnamefont{R.}~\bibnamefont{Bolgiano}},
  \bibinfo{journal}{Journal of Geophysical Research}
  \textbf{\bibinfo{volume}{67}}, \bibinfo{pages}{3015} (\bibinfo{year}{1959}).

\bibitem[{\citenamefont{Obukhov}(1959)}]{obukhov:convecpheno}
\bibinfo{author}{\bibfnamefont{A.~M.} \bibnamefont{Obukhov}},
  \bibinfo{journal}{Doklady Akademiia Nauk {SSSR}}
  \textbf{\bibinfo{volume}{125}}, \bibinfo{pages}{1246} (\bibinfo{year}{1959}).

\bibitem[{\citenamefont{Grappin{, U. Frisch, J. L\'eorat und A.
  Pouquet}}(1982)}]{grappin_etal:zpzmspectra}
\bibinfo{author}{\bibfnamefont{R.}~\bibnamefont{Grappin{, U. Frisch, J.
  L\'eorat und A. Pouquet}}}, \bibinfo{journal}{Astronomy and Astrophysics}
  \textbf{\bibinfo{volume}{105}}, \bibinfo{pages}{6} (\bibinfo{year}{1982}).

\bibitem[{\citenamefont{Grappin et~al.}(1983)\citenamefont{Grappin, Pouquet,
  and L\'eorat}}]{grappin_pouquet_leorat:correl}
\bibinfo{author}{\bibfnamefont{R.}~\bibnamefont{Grappin}},
  \bibinfo{author}{\bibfnamefont{A.}~\bibnamefont{Pouquet}}, \bibnamefont{and}
  \bibinfo{author}{\bibfnamefont{J.}~\bibnamefont{L\'eorat}},
  \bibinfo{journal}{Astronomy and Astrophysics} \textbf{\bibinfo{volume}{126}},
  \bibinfo{pages}{51} (\bibinfo{year}{1983}).

\bibitem[{\citenamefont{Dobrowolny et~al.}(1980)\citenamefont{Dobrowolny,
  Mangeney, and Veltri}}]{dobrowolny:ik}
\bibinfo{author}{\bibfnamefont{M.}~\bibnamefont{Dobrowolny}},
  \bibinfo{author}{\bibfnamefont{A.}~\bibnamefont{Mangeney}}, \bibnamefont{and}
  \bibinfo{author}{\bibfnamefont{P.}~\bibnamefont{Veltri}},
  \bibinfo{journal}{Physical Review Letters} \textbf{\bibinfo{volume}{45}},
  \bibinfo{pages}{144} (\bibinfo{year}{1980}).

\end{thebibliography}
\newcommand{\nop}[1]{}

\end{document}